\documentclass[a4paper,11pt]{article}
\pdfoutput=1 

\usepackage{jcappub}

\usepackage[T1]{fontenc} 

\usepackage{tabularx}

\title{\boldmath Cosmic ray mass composition at the \textit{knee} using azimuthal fluctuations of air shower particles detected at ground by the KASCADE experiment}

 \author{N. Arsene}
 \affiliation{Institute of Space Science,\\P.O.Box MG-23, Ro 077125 Bucharest-Magurele, Romania}

\emailAdd{nicusorarsene@spacescience.ro}

\abstract{The presence of hadronic sub-showers causes azimuthal non-uniformity in the particle distributions on the ground in vertical air showers. The $LCm$ parameter, which quantifies the non-uniformity of the signal recorded in detectors located at a given distance on a ring around the shower axis, has been successfully introduced as a gamma/hadron discriminator at PeV energies \cite{Conceicao:2022lkc}. In this work, we demonstrate that the $LCm$ parameter can effectively serve as a mass composition discriminator in experiments that employ a compact array of detectors, like KASCADE. 
We reconstruct the $LCm$ parameter distributions in the energy range $\lg(E/\rm eV) = [15.0 \text{ - } 16.0]$ using measurements from the KASCADE experiment, with intervals of $\lg(E/\rm eV) = 0.2$, which are then fitted with MC templates for five primary nuclei species p, He, C, Si, and Fe considering three hadronic interaction models: QGSjet-II-04, EPOS-LHC and SIBYLL 2.3d. 
We find that the $LCm$ parameter exhibits minimal dependence on the specific hadronic interaction model considered.
The reconstructed fractions of individual species demonstrate a linear decrease in the abundance of protons and He nuclei with increasing energy, while the heavier components become prevalent above the \textit{knee} as predicted by all three hadronic interaction models. Our findings indicate that the abundance of particle types as a function of energy aligns with different astrophysical models that link the \textit{knee} to the acceleration and propagation of cosmic rays within the Galaxy. Furthermore, they also demonstrate excellent agreement with three more recent data-driven astrophysical models. These findings suggest that the $LCm$ parameter could be a valuable tool for forthcoming measurements of the LHAASO experiment to enhance our knowledge about the origin and acceleration mechanisms of cosmic rays.}

\begin{document}
\maketitle
\flushbottom

\section{Introduction}
\label{sec:intro}

Various ground-based experiments have measured the energy spectrum of cosmic rays for several decades. To understand their origin and acceleration mechanisms, it is essential to accurately determine the mass composition of the cosmic rays, as well as their arrival directions, and correlate these measurements with a precise picture of the galactic/extragalactic magnetic fields.
The experimental measurements reveal some remarkable structures in the flux of the primary cosmic rays ($dN/dE \propto E^\gamma$): a change of the spectral index $\gamma$ from -2.7 to -3.1 around $E \sim 4 \times 10^{15}$ eV (the \textit{knee}) \cite{KASCADE:2005ynk,Takeda:2002at,Nagano:1984db,Ogio:2004sc,Fowler:2000si,Glasmacher:1999id,Swordy:1999um,EAS-TOP:1998aye}, a change from $\gamma = - 2.95$ to $\gamma = - 3.24$ at  $E \sim 8 \times 10^{16}$ eV (the \textit{second knee}) \cite{KASCADEGrande:2011kpw,IceCube:2019hmk,TelescopeArray:2018bya} and a flattening in the spectrum at $E \sim 5 \times 10^{18}$ eV (the \textit{ankle}) where the $\gamma$ index changes from -3.29 to -2.51 \cite{Deligny:2020gzq,PierreAuger:2021hun,PierreAuger:2020kuy,TelescopeArray:2015dcv}.

Different scenarios have been proposed to explain the appearance of the \textit{knee} in the energy spectrum. The most widely considered suggest that the \textit{knee} is due to the maximum energy that can be reached by cosmic rays accelerated in supernova remnants e.g. \cite{Stanev:1993tx, Sveshnikova:2003sa, Plaga:2001dk}, or it is due to the effects of cosmic ray propagation through the interstellar medium e.g. \cite{Roulet:2003rr, Swordy95}.
Another class of models explain the \textit{knee} as a consequence of the interaction of cosmic rays with the photon fields \cite{Karakula:1993he}. A detailed description of these models compared with direct and indirect measurements of cosmic rays properties around the \textit{knee} is given in \cite{Hoerandel:2004gv}.

Measurements of the mass composition around the \textit{knee} indicate an increase in the mean logaritmic mass ($\ln A$) with increasing energy of the primary cosmic rays, which suggests that individual elements may be undergoing sequential cut-offs \cite{Horandel:2005at,KASCADE:2004vrv,Haungs:2002ck,KASCADE:2005ynk}. However, an accurate interpretation of the results is limited by discrepancies between the considered hadronic interaction models.

In this work, we present a new mass composition reconstruction method that utilizes the non-uniformity of the signal recorded in detectors located at a fixed distance on a ring around the shower axis at ground level. Originally introduced as a gamma/hadron discrimination technique at PeV energies \cite{Conceicao:2022lkc}, we demonstrate that this method shows minimal dependence on the specific hadronic interaction model and can effectively serve as mass composition discriminator in cosmic ray experiments with a relatively compact array of detectors, such as KASCADE \cite{KASCADE:2003swk}.

The paper is organized as follows: in Section \ref{sec2} we introduce the $LCm$ parameter which is used as mass discriminator at PeV energies, in Section \ref{secMC} we describe how the Monte Carlo (MC) simulations were obtained, in Section \ref{secKASCADE} we reconstruct the distributions of $LCm$ from KASCADE data \cite{Haungs:2018xpw}, in Section \ref{sec_syst} we estimate the systematic uncertainties of the fitted fractions,  in Section \ref{secMassComp} we perform the mass composition reconstruction in the energy range $\lg(E/\rm eV) = [15.0 \text{ - } 16.0]$ on the basis of three hadronic interaction models and compare the results with predictions of different astrophysical models of the \textit{knee}. Section \ref{secConcl} concludes the paper.

\section{The $LCm$ observable}
\label{sec2}

The stochastic nature of the extensive air showers (EAS) development in Earth's atmosphere produces non-uniformity in particle densities at ground level. In vertical showers, we would expect similar densities of secondary particles at a fixed distance around the shower core. Neglecting the influence of the geomagnetic field, the presence of hadronic sub-showers can distort this symmetry. At the same energy, a gamma-induced shower will present a less complex structure of particle densities on the ground than a proton-induced shower. This is because the development of gamma showers is mainly driven by electromagnetic interactions. In the same context, we expect larger fluctuations in the azimuthal signal in proton-induced showers compared to iron showers due to the larger fluctuations in primary interaction heights in the atmosphere. To account for such non-uniformity in the signal induced in detectors at a given distance from the shower axis, the $LCm$ parameter was introduced as a promising gamma/hadron discriminator \cite{Conceicao:2022lkc}.

The $LCm$ parameter is computed as $LCm = \log(C_k)$, where the $C_{k}$ variable is defined for each radial ring $k$ around the shower core at ground as \cite{Conceicao:2022lkc}:

\begin{equation}
C_{k} =\frac{2}{n_{k}(n_{k}-1)} 
\frac{1}{\left<S_{k}\right>}\sum_{i=1}^{n_{k}-1}\sum_{j=i+1}^{n_{k}}(S_{ik}-S_{jk})^{2} ,   
\label{eq:CK}
\end{equation}
where $n_{k}$ represents the number of detectors in ring $k$, $\left<S_{k}\right>$ is the mean signal registered in the detectors of the ring $k$, while $S_{ik}$ and $S_{jk}$ stand for the signals in the detectors  $i$ and $j$ of the ring $k$, respectively.
The term $\frac{2}{n_{k}(n_{k}-1)}$ accounts for the inverse of the number of two-combinations for $n_k$ detectors, $\binom{n_k}{2}$. 
A greater non-uniformity in the azimuthal distribution of the signal results in an increased value of the $C_k$ variable. In contrast, a perfectly uniform azimuthal distribution results in $C_k$ approaching zero.

The value of $LCm$ parameter depends on the primary energy and the fill factor (FF) of the detectors array. 
For a fixed primary energy, the value of $LCm$ parameter as a function of radial range is roughly constant above $\gtrsim 40$ m up to $1000$ m. In this analysis we reconstruct the $LCm$ parameter for both data and simulations in the energy range $\lg(E/\rm eV) = [15.0 \text{ - } 16.0]$ in intervals of $\lg(E/\rm eV) = 0.2$, considering the radial range $r_k = [100 \text{ - } 110]$ m. The following section will explain the methodology used to generate simulation sets for five primary species, namely p, He, C, Si, and Fe, using various hadronic interaction models. The simulation sets were then used to construct $LCm$ distributions in each energy interval, which are then used in determining the mass composition of cosmic rays using data collected by the KASCADE experiment.

\section{$LCm$ distributions from Monte Carlo simulations of KASCADE array}
\label{secMC}

\begin{figure}[tbp]
\centering 
\includegraphics[width=.6\textwidth]{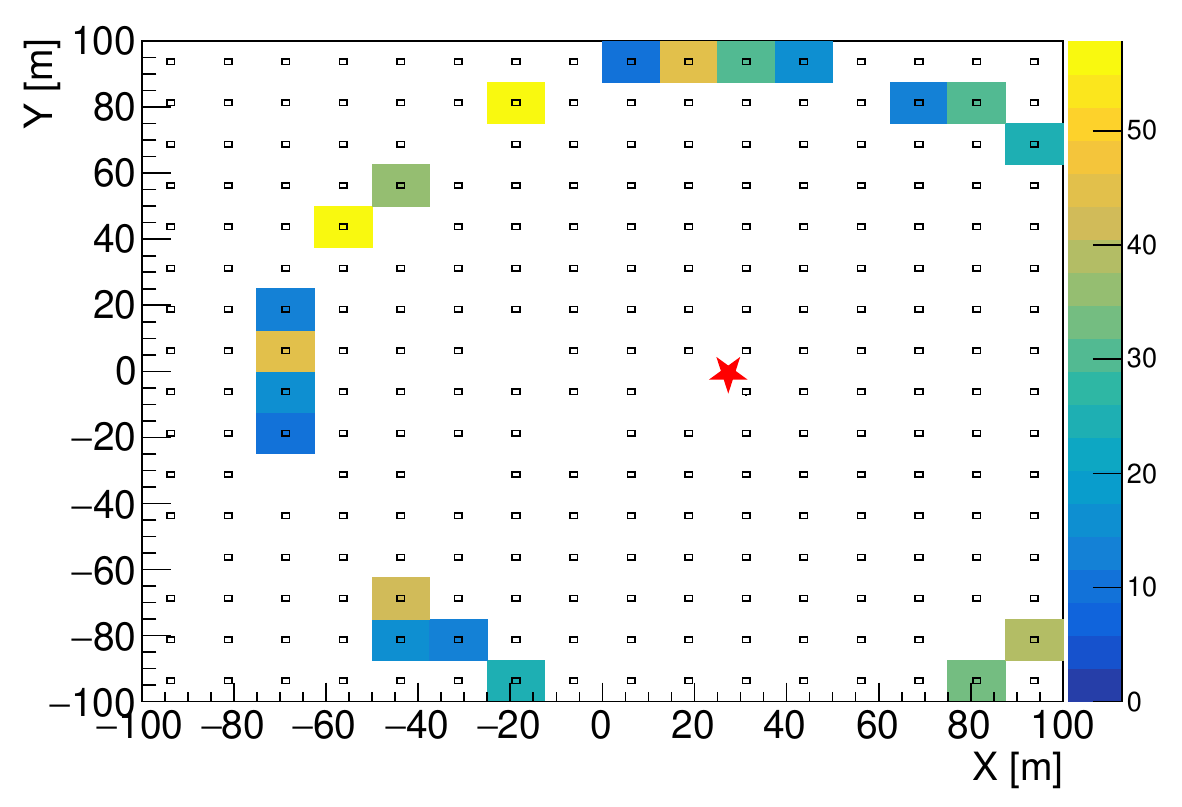}
\caption{\label{fig_array} Visualization of the 252-detector array of the KASCADE experiment. The color gradient represents the energy deposited in [MeV] by the electromagnetic component in $e/\gamma$-detectors located in the radial range $r_k = [100 \text{ - } 110]$ m from the shower core position indicated by the red star. The reconstructed parameters of this event with the internal KASCADE counting number 302724 are: the primary energy $\lg(E/\rm eV) = 15.6$, the zenith angle of the shower axis $\theta = 12.1^{\circ}$ and the azimuth angle $\phi = 53.7^{\circ}$.}
\end{figure}

The KASCADE experiment utilized a detector array consisting of 252 stations arranged on a rectangular grid with a 13 m spacing covering an area of $200 \times 200$ m$^{2}$, being able to reconstruct the properties of extensive air showers with energies in the range $\lg (E /\rm eV) = [14 \text{ - } 17]$. The stations were equipped with both shielded and unshielded detectors, allowing for the simultaneous and independent measurement of the electromagnetic and muonic components from extensive air showers. Above a 3 MeV threshold, the liquid scintillation counters located above the shielding were used to detect the charged component of the shower's secondary particles. The muonic component was measured by utilizing $3.2$ m$^{2}$ of plastic scintillators that were placed beneath lead and iron absorber sheets \cite{KASCADE:2003swk}.
In Figure \ref{fig_array} we represent the layout of 252-detector array of the KASCADE experiment together with the energy deposited by the electromagnetic component in $e/\gamma$-detectors located at $r_k = [100 \text{ - } 110]$ m from the shower axis, from the event with the internal counting number 302724, with the primary energy $\lg(E/\rm eV) = 15.6$, zenith angle of the shower axis $\theta = 12.1^{\circ}$ and the azimuth angle $\phi = 53.7^{\circ}$.

The MC simulations of extensive air showers for KASCADE experiment is a three-step procedure \cite{Haungs:2018xpw, Wochele-sim-kcdc}. The development of showers in the atmosphere is simulated with CORSIKA \cite{corsika,corsika1}, the signal / energy deposit in the KASCADE detectors is simulated with CRES package based on GEANT3 \cite{Brun:1987ma} while the reconstruction of the air shower observables like primary energy, arrival direction, the number of electrons / muons / hadrons and so on, is performed with the KRETA package.
It should be noted that the KRETA package, which follows the same procedures, is utilized for reconstructing the events that were recorded by KASCADE experiment.

Three hadronic interaction models at high energies were considered: EPOS-LHC \cite{Pierog:2013ria}, QGSjet-II-04 \cite{Ostapchenko:2004ss} and SIBYLL 2.3d \cite{Riehn:2019jet}, while the low energy hadronic interactions ($E_{\mathrm{lab}} < 200$ GeV) were modeled with FLUKA \cite{Ferrari:898301}. Five species of primary particles were simulated (p, He, C, Si, and Fe) in the energy range $\lg(E/\mathrm{eV}) = [15.0 \text{ - } 16.0]$ with intervals of $\lg(E/\mathrm{eV}) = 0.2$ considering the energy spectral index $\gamma = -2.7$. The zenith angles of the shower axis were isotropically sampled from the range $\theta = [0^{\circ} \text{ - } 20^{\circ}]$ and the azimuthal distribution was uniformly sampled in the range $\phi = [0^{\circ} \text{ - } 360^{\circ}]$. The statistics in the simulation sets is of the order of $10^{3} \text{ - } 10^{4}$ events / primary species / model in each energy interval.

\begin{figure}[tbp]
\centering 
\includegraphics[width=.49\textwidth]{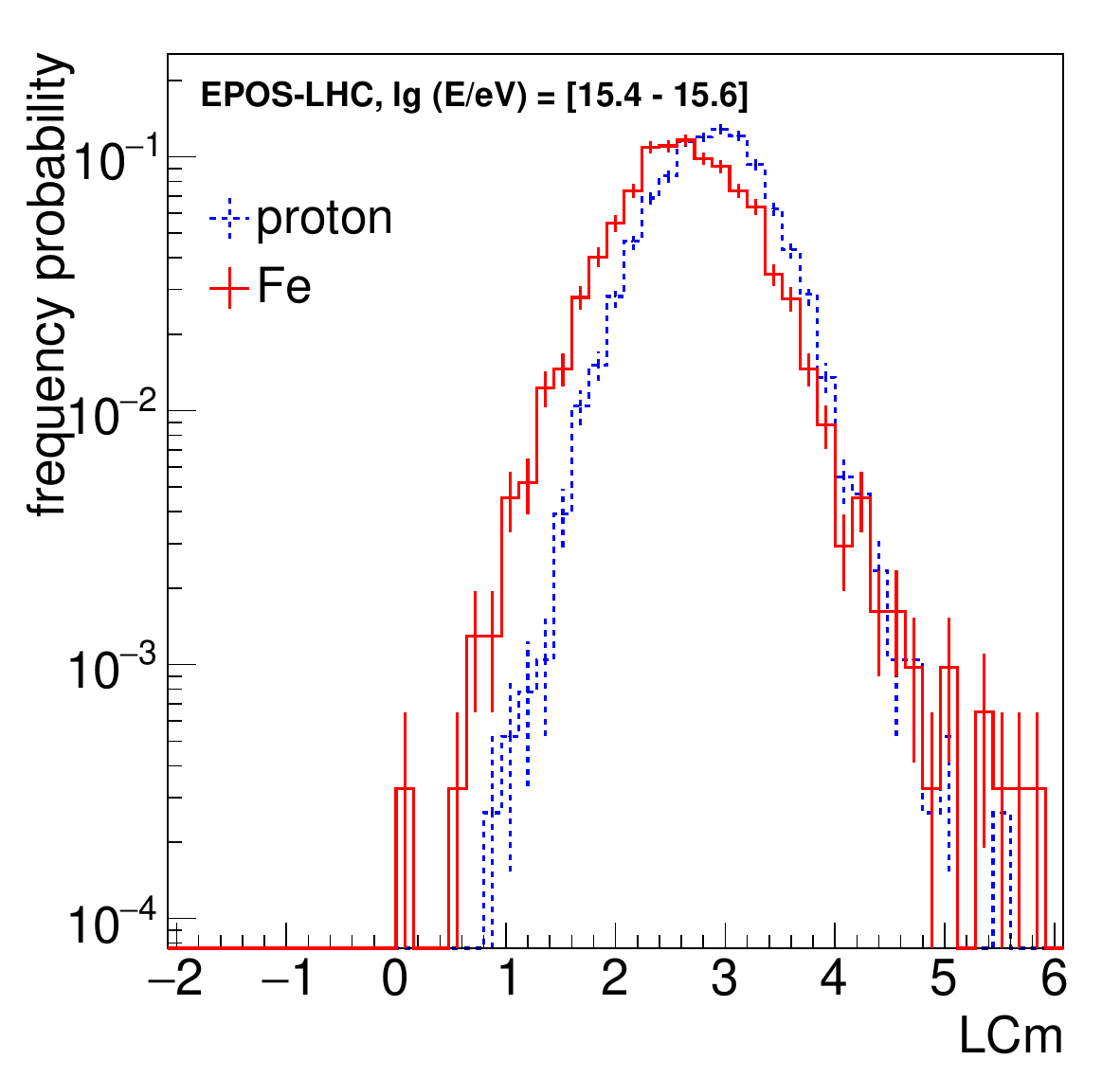}
\caption{\label{fig_p_vs_fe} The $LCm$ distributions for proton induced showers \textit{(dashed blue)} and Fe induced showers \textit{(continuous red)} in the energy interval $\lg(E/\mathrm{eV}) = [15.4 \text{ - } 15.6]$ on the basis of EPOS-LHC and FLUKA as hadronic interaction models at high and low energies, respectively.}
\end{figure}

For each primary species, hadronic interaction model and energy interval we build $LCm$ distributions using Equation \ref{eq:CK} where the signal $S_i$ represents the energy deposit of the electromagnetic component in the $"i"$ $e/\gamma$-detector from the radial interval $r_k = [100 \text{ - } 110]$ m around the shower core position. This quantity is separately and independent measured by the muonic and hadronic energy deposited in KASCADE detectors. In Figure \ref{fig_p_vs_fe} are represented the $LCm$ distributions reconstructed for proton and Fe induced showers based on EPOS-LHC hadronic interaction model in the energy interval $\lg(E/\mathrm{eV}) = [15.4 \text{ - } 15.6]$. The shapes of the two distributions are distinct, as well as their mean and dispersion values. The larger $LCm$ values for proton induced showers are due to the larger fluctuations in the heights of the first interaction points in the atmosphere.
\begin{figure}[tbp]
\centering 
\includegraphics[width=.49\textwidth]{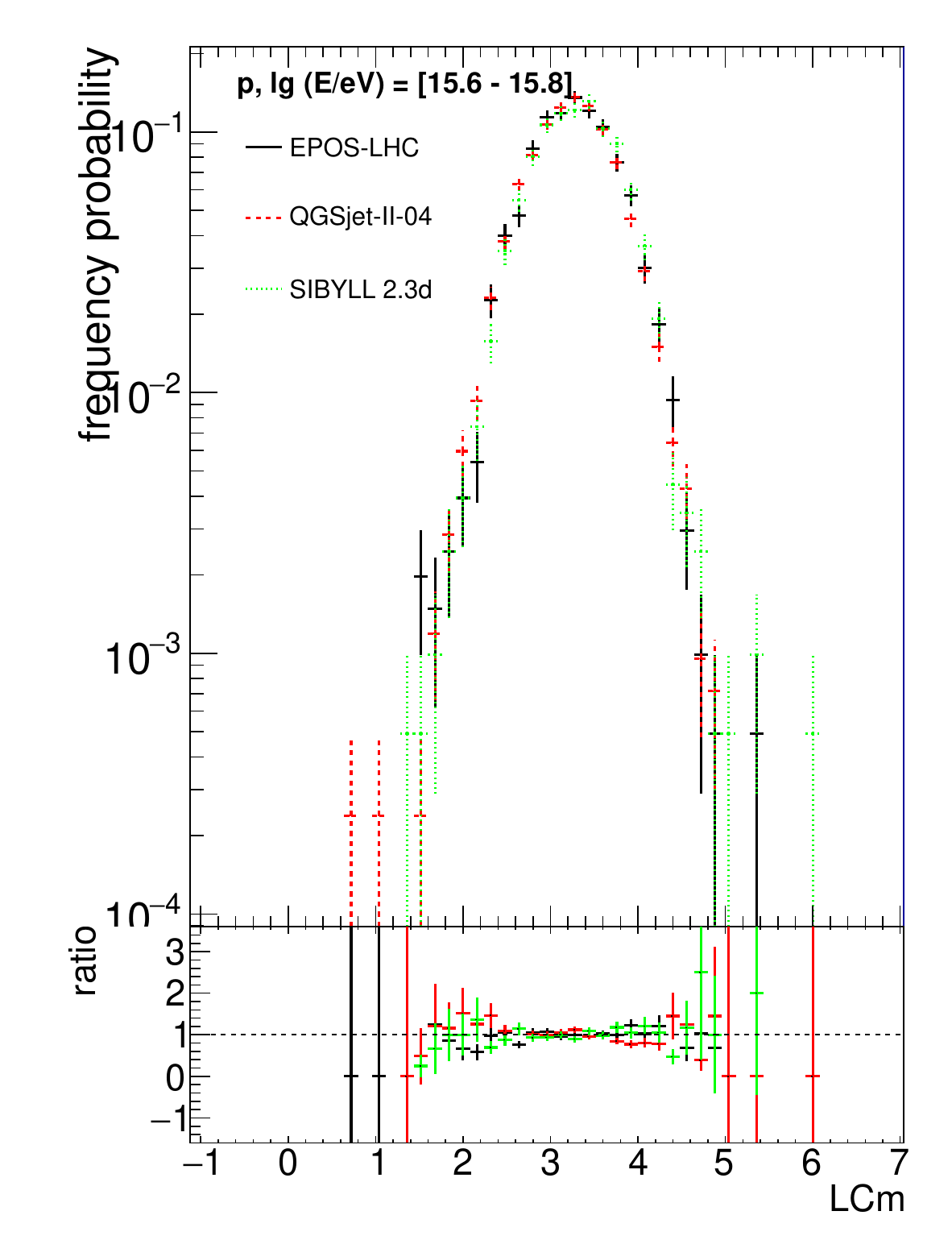}
\hfill
\includegraphics[width=.49\textwidth,origin=c]{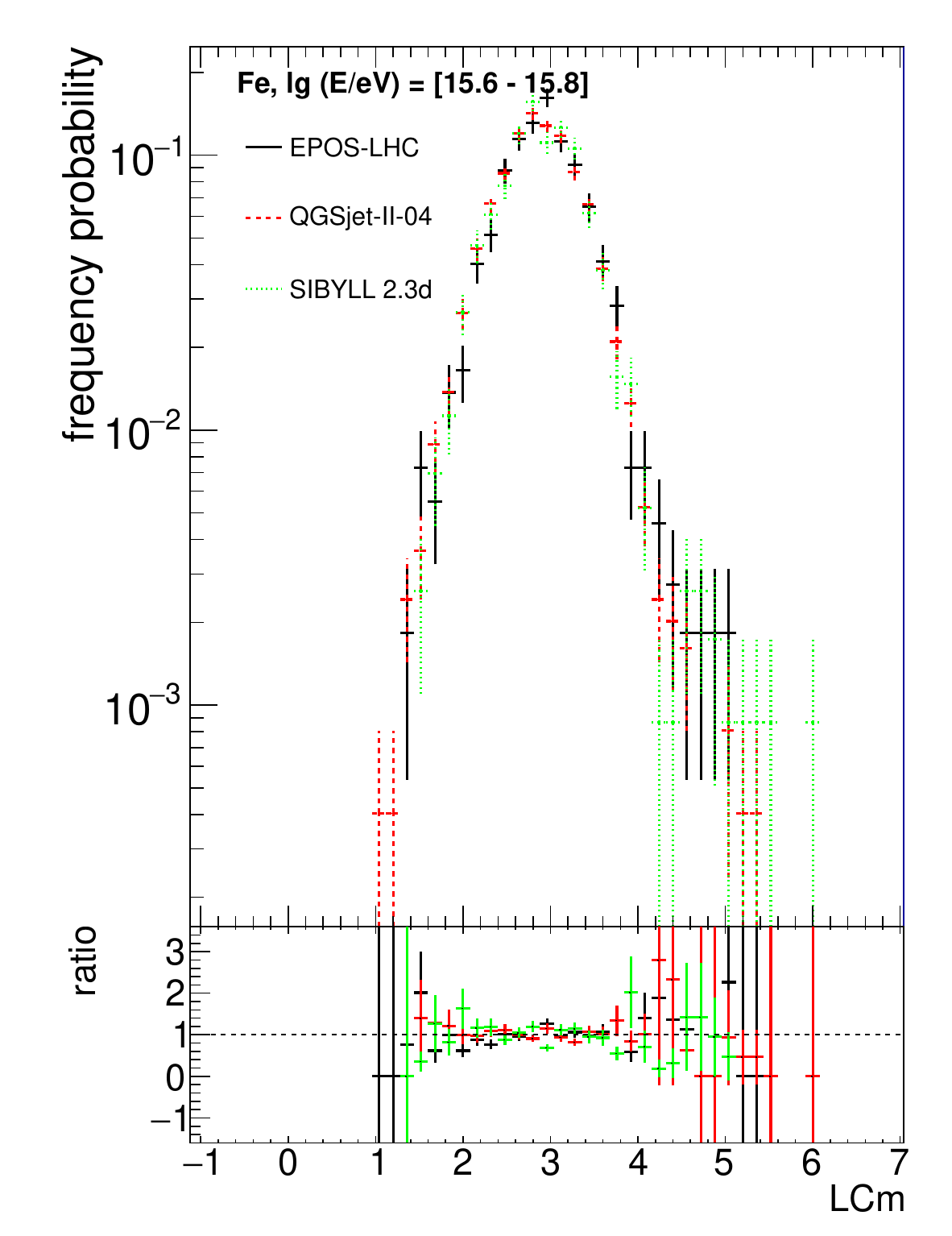}
\caption{\label{fig_all_models} The $LCm$ distributions for proton \textit{(left)} and Fe \textit{(right)} in the energy interval $\lg(E/\mathrm{eV}) = [15.6 \text{ - } 15.8]$ as predicted by all three hadronic interaction models considered in this work. The ratio of each pair of two distributions is displayed at the bottom of each plot.}
\end{figure}

\begin{figure}[!htb]
\minipage{0.31\textwidth}
  \includegraphics[width=\linewidth]{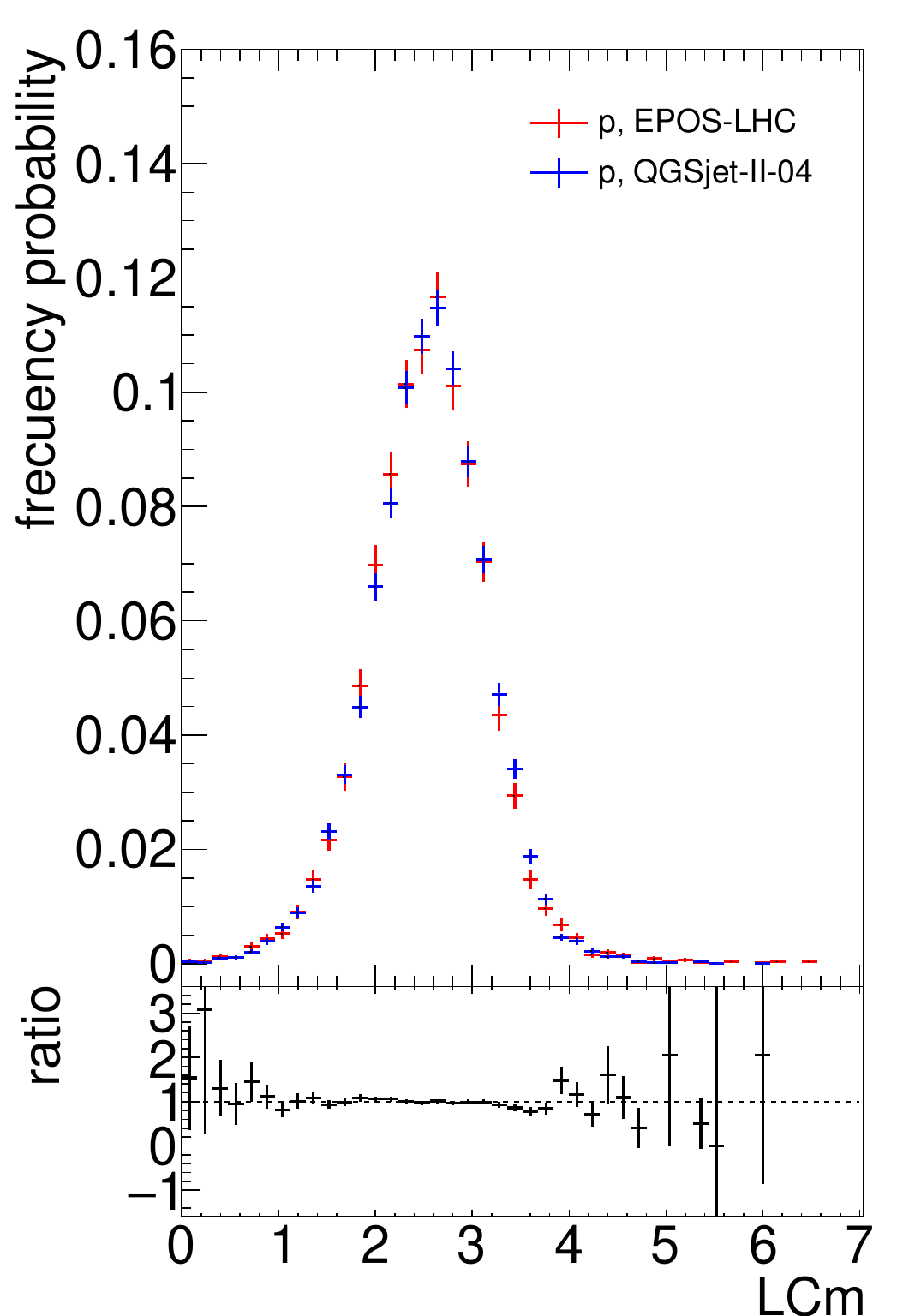}
\endminipage\hfill
\minipage{0.31\textwidth}
  \includegraphics[width=\linewidth]{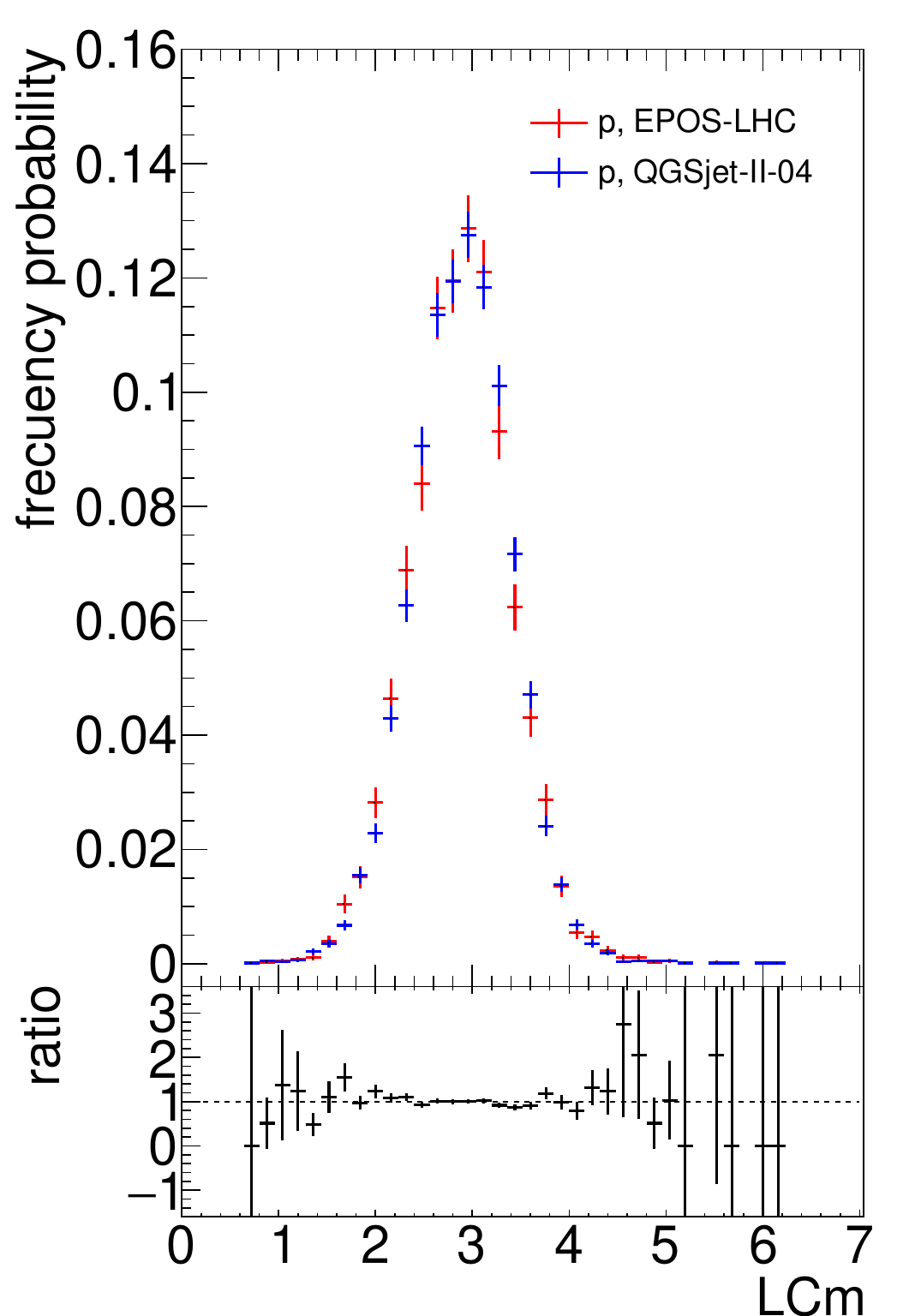}
\endminipage\hfill
\minipage{0.31\textwidth}%
  \includegraphics[width=\linewidth]{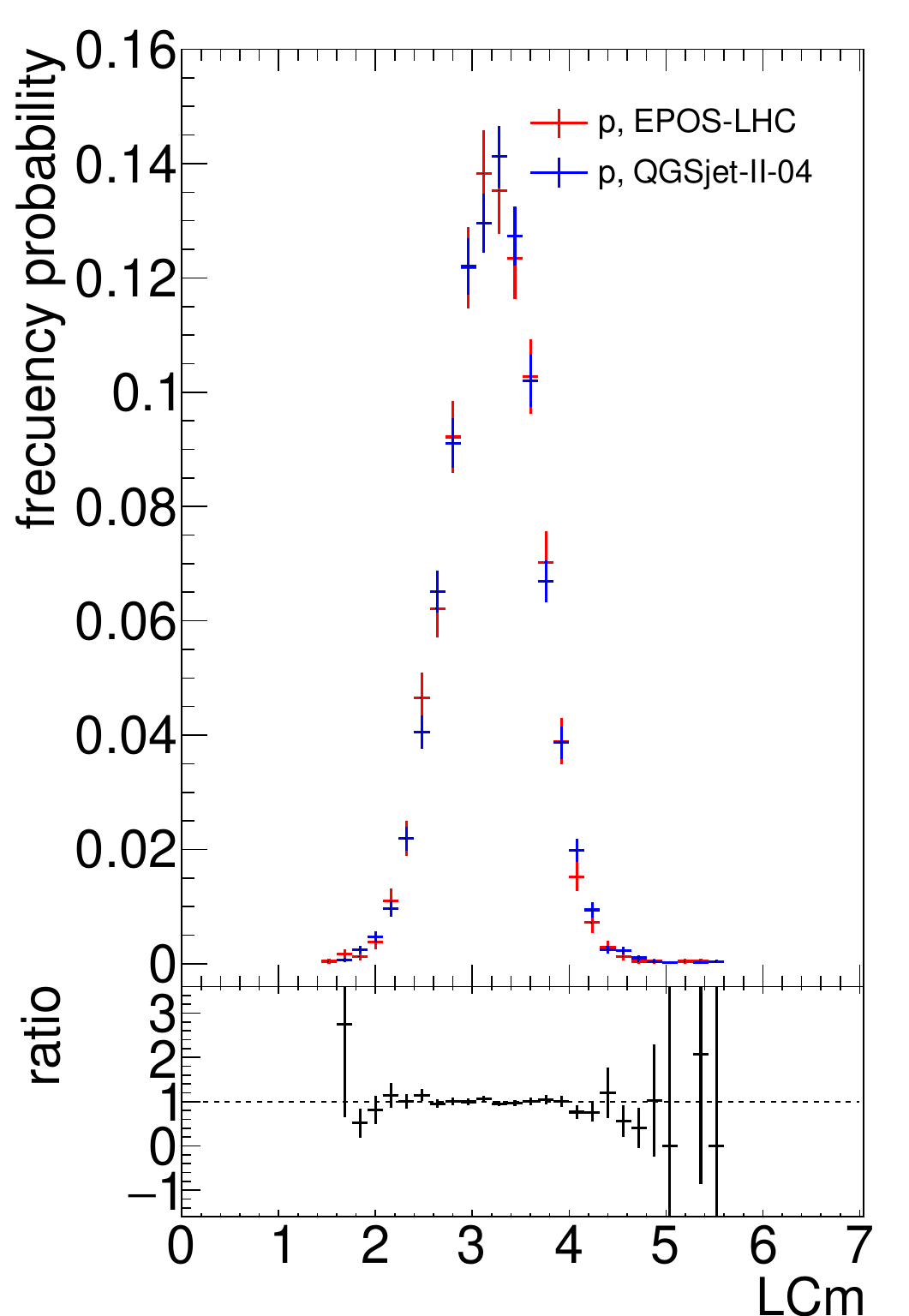}
\endminipage
  \caption{The $LCm$ distributions obtained for proton-induced showers with EPOS-LHC \textit{(red)} and QGSjet-II-04 \textit{(blue)} in three energy intervals $\lg(E/\mathrm{eV}) = [15.2 \text{ - } 15.4]$ \textit{(left)}, $\lg(E/\mathrm{eV}) = [15.4 \text{ - } 15.6]$ \textit{(middle)} and $\lg(E/\mathrm{eV}) = [15.6 \text{ - } 15.8]$ \textit{(right)}. The ratio of the distributions is displayed on the bottom of each plot.}\label{fig_ratios}
\end{figure}

A very important aspect of the $LCm$ parameter is that it exhibits minimal dependence on the specific hadronic interaction model used in the simulation process of extensive air showers. Therefore, using experimental data obtained from the KASCADE experiment, the mass composition of cosmic rays can be determined in a minimally model-dependent way. In Figure \ref{fig_all_models} we represent the $LCm$ distributions in the energy bin $\lg (E/\mathrm{eV}) = [15.6 \text{ - } 15.8]$ for proton and Fe induced showers, as predicted by all hadronic interaction models considered in this analysis. The lower plots quantify the ratio between each pair of two distributions. It is evident that the models show a remarkable level of agreement. The same degree of compatibility is obtained on the entire energy spectrum analyzed in this paper $\lg(E/\mathrm{eV}) = [15.0 \text{ - } 16.0]$. Figure \ref{fig_ratios} provides an example of the quantitative agreement between $LCm$ distributions that are predicted by different models. We compare the $LCm$ distributions obtained for proton-induced showers with EPOS-LHC and QGSjet-II-04 in three energy intervals: $\lg(E/\mathrm{eV}) = [15.2 \text{ - } 15.4]$, $\lg(E/\mathrm{eV}) = [15.4 \text{ - } 15.6]$ and $\lg(E/\mathrm{eV}) = [15.6 \text{ - } 15.8]$. The comparison is made by computing the ratio of the two distributions. The values of the \textit{ratio} parameter displayed at the bottom of the plots are very close to 1 throughout the distribution, within the limits of statistical uncertainties. 
Using these distributions, we will perform a mass composition analysis by fitting the experimental $LCm$ distributions for each energy interval with MC templates for five primary species (p, He, C, Si and Fe) in Section \ref{secMassComp}.

\section{$LCm$ distributions from KASCADE data}
\label{secKASCADE}

Our analysis focused on the experimental data collected by KASCADE from 1996 until the end of its final operation in 2013 \cite{Haungs:2018xpw}. Specifically, we considered data within the energy range $\lg(E/\mathrm{eV}) = [15.0 \text{ - } 16.0]$ and restricted our analysis to vertical showers, i.e., those with a zenith angle $\theta$ in the range of $[0^{\circ} \text{ - } 20^{\circ}]$. This was done to prevent potential biases in the azimuthal distribution of the signal resulting from attenuation and geometric effects \cite{Sima:2011zz}. 

In the initial analysis using the KRETA reconstruction program, the collaboration applied the "Data Selection Cuts KASCADE" to the data \cite{Wochele-kcdc}. These cuts are deemed essential by the collaboration to ensure a consistent level of quality in the published data sets. These cuts consist in: a successful reconstruction process of the event in the KASCADE array Processor, the array has produced a trigger signal, no more than 2 out of the 16 array clusters to be missing, 
the reconstructed shower core is within a 91 m radius around the centre of the array of detectors, the lateral shower shape parameter $s$ has to be in the range $s = [0.1 \text{ - } 1.48]$, the zenith angle of the shower axis is $\theta < 60^{\circ}$, the number of electrons $N_e > 100$ and the number of muons $N_\mu > 100$. More stringent cuts are highly recommended by the KASCADE collaboration, which we took into account both in data and simulation reconstruction procedures. These cuts impose $\theta < 42^{\circ}$, $s = [0.6 \text{ - } 1.3]$ and $N_e > 100000$. Important to mention that the $e/\gamma$-detectors of the KASCADE array reach the full efficiency at $\lg (N_e) > 4.25$, therefore, no biases are expected in the reconstruction process of the $LCm$ parameter.

\begin{figure}[tbp]
\centering 
\includegraphics[width=1\textwidth]{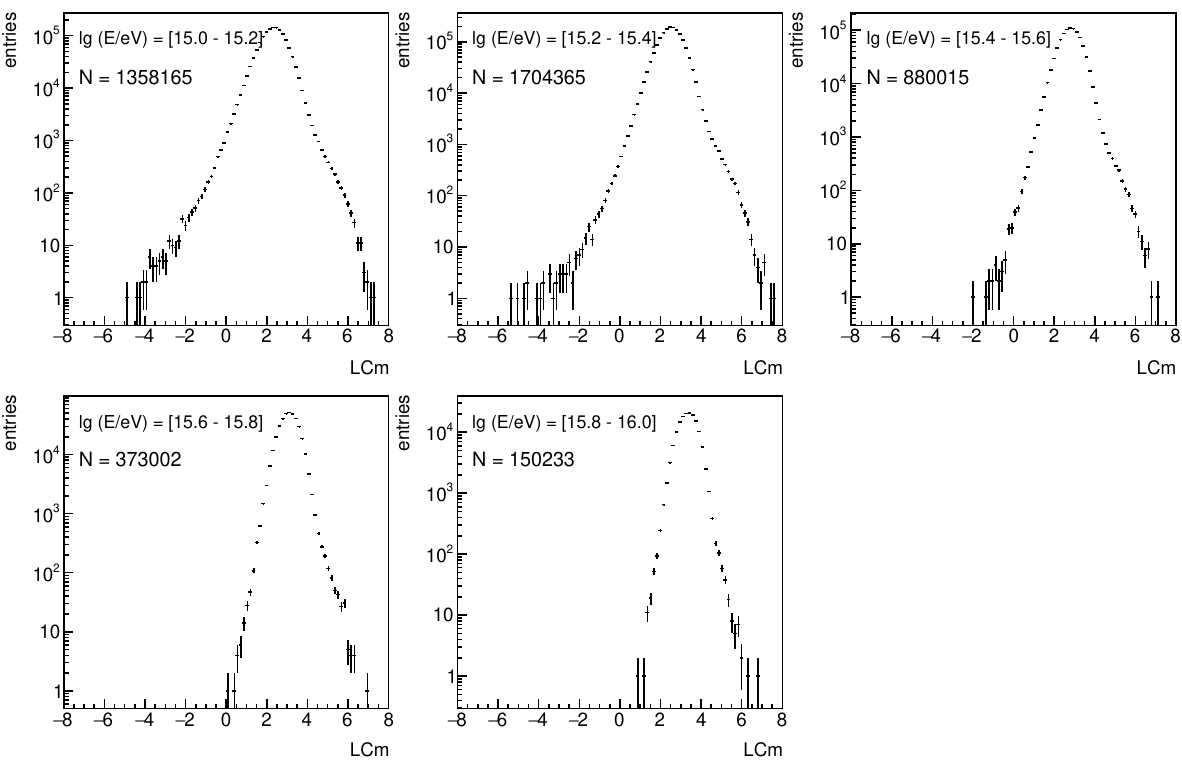}
\caption{\label{fig_LCm_data} The $LCm$ distributions reconstructed from KASCADE data in each energy interval in the range $\lg(E/\mathrm{eV}) = [15.0 \text{ - } 16.0]$. The number of events $N$ that survived all applied cuts is indicated on each plot.}
\end{figure}

In Figure \ref{fig_LCm_data} we present the $LCm$ distributions in each energy interval within the range of $\lg(E/\mathrm{eV}) = [15.0 \text{ - } 16.0]$ reconstructed from KASCADE data. Note that these distributions represent the \textit{raw data}, not yet  corrected for \textit{bin-to-bin migration} effect as described in Section \ref{sec_syst}. We have applied the aforementioned cuts and only the events that have passed the reconstruction procedure are included. The number of surviving events $N$ is indicated on the plots for each energy interval of $\lg(E/\mathrm{eV}) = 0.2$.
As can be seen, the mean of the $LCm$ distributions increases with increasing primary energy, and the spread of the distributions become narrower as the primary energy increases. In the upcoming section, we will use the MC predictions derived in Section \ref{secMC} to fit the experimental $LCm$ distributions. This will enable us to investigate the dependence of the individual abundance of primary nuclei on primary energy within the energy range of $\lg(E/\mathrm{eV}) = [15.0 \text{ - } 16.0]$. 

\section{Systematic uncertainties}
\label{sec_syst}
To perform a mass composition analysis by fitting the experimental data recorded in the KASCADE experiment with MC templates obtained from simulations, we estimated the systematic uncertainties due to the primary energy reconstruction and the dependencies on MC simulations.

We performed an estimation of systematic errors in the reconstruction of primary energy using true energies ($E^{true}$) obtained from CORSIKA simulations and reconstructed energies ($E^{rec}$) obtained from detector response simulations using the CRES package. Cascade parameters were reconstructed using the KRETA package. To evaluate the systematic uncertainty in energy reconstruction, we divided the quantity $(E^{true} - E^{rec})/E^{true}$ into energy intervals of $\lg(E/\rm eV) = 0.1$. We averaged these values over the three hadronic interaction models and assigned equal weights to all five species of primary particles. The standard deviation of each distribution represented the estimated systematic uncertainty.

The bias values, which represent the mean values of each $(E^{true} - E^{rec})/E^{true}$ distribution, ranged from 1\% to 3\%. The systematic uncertainties are determined to lie within 20\% and 29\%.
A first measure to avoid strong correlations between adjacent energy intervals, was to bin the data into intervals of $\lg(E/\rm eV) = 0.2$.

Another step involved addressing the \textit{bin-to-bin migration} effect caused by systematic errors during the reconstruction of primary energy. For each hadronic interaction model, primary particle species and energy, we generated distributions of $(E^{true} - E^{rec})$ through simulations. These distributions provided the probability of an event being reconstructed with a certain energy in one energy bin "$i$" while actually belonging to energy bin "$j$".
To account for this effect, we re-binned the experimental data into energy intervals, taking into account the modified energy values based on the probability distributions of \textit{bin-to-bin migration} for the three hadronic interaction models (QGSjet-II-04, EPOS-LHC, and SIBYLL 2.3d). As a result, we obtained three sets of experimentally corrected data for \textit{bin-to-bin migration} corresponding to each model.

Considering that the distributions of $(E^{true} - E^{rec})$ within a particular energy interval slightly depend on the nature of the primary particle, we considered various scenarios when correcting the energies of the experimental data. These scenarios included predominantly light nuclei (proton + He), predominantly heavy nuclei, or a combination involving also intermediate masses. After careful analysis, we determined that the optimal approach is to consider an equal proportion of light and heavy masses. This choice aims to maximize the goodness-of-fit values when fitting the data with MC templates. Note that when considering the light elements scenario or the heavy elements scenario, the results of the fitted fractions of individual species on experimental data do not change in a significant way.

Next, we estimated the sensitivity, bias and systematic errors of the method due to MC mismodeling. 
To evaluate the sensitivity of this method, we conducted the following test. Firstly, we utilized individual $LCm$ values obtained from simulations based on a specific hadronic interaction model. By considering known random abundances of the five primary species (p, He, C, Si, and Fe), we constructed $LCm$ distributions for each energy bin. To ensure comprehensive coverage, we generated a large number of such distributions ($\sim$10000) encompassing all feasible combinations.

\begin{figure}
  \includegraphics[width=\linewidth]{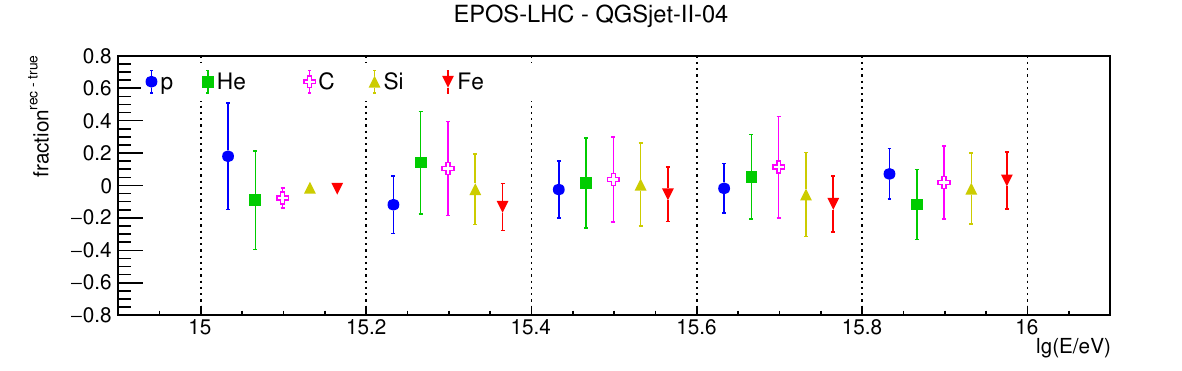}
  \includegraphics[width=\linewidth]{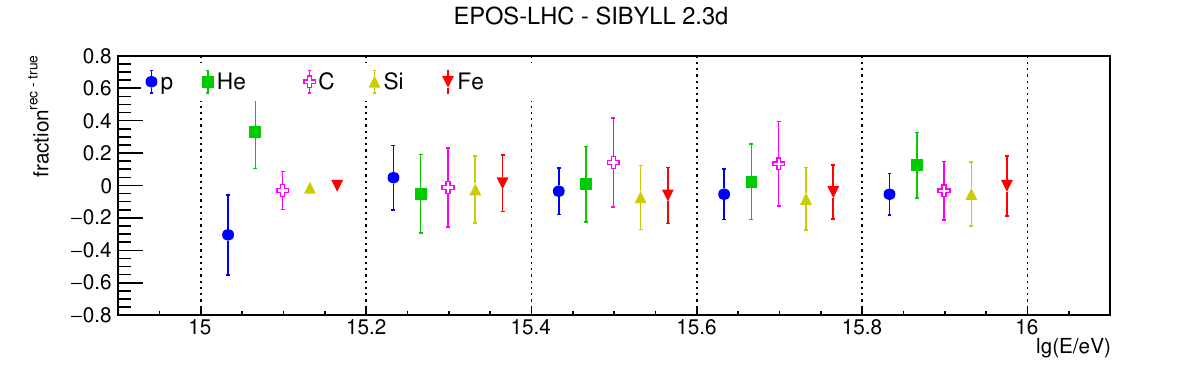}  
    \caption{The values of $fraction^{rec - true}$ as a function of energy for the case in which the $LCm$ distributions are constructed on the basis of EPOS-LHC model and fitted with MC templates predicted by QGSjet-II-04 \textit{(up)} and SIBYLL 2.3d \textit{(down)}.}
  \label{fig_epos}
\end{figure}

Subsequently, we fitted these $LCm$ distributions with five MC templates (p, He, C, Si, and Fe) predicted by another hadronic interaction model. We then compared the reconstructed fractions to the true fractions of each species within each energy interval, calculating the differences ($fraction^{rec - true}$). 
Figure \ref{fig_epos} depicts the $fraction^{rec - true}$ values when constructing $LCm$ distributions from EPOS-LHC simulations and fitting them with MC templates from the QGSjet-II-04 and SIBYLL 2.3d hadronic interaction models.
This procedure was repeated for all possible permutations of the three hadronic interaction models considered in this analysis. The error bars in the figures represent the standard deviations of the $fraction^{rec - true}$ distributions within each energy interval.

These uncertainties, denoted as $\sigma_{syst}^1$, are considered as one source of systematic uncertainties when applying the proposed method to KASCADE data. We performed the same test for all permutations of the three hadronic interaction models.

\begin{figure}
  \includegraphics[width=\linewidth]{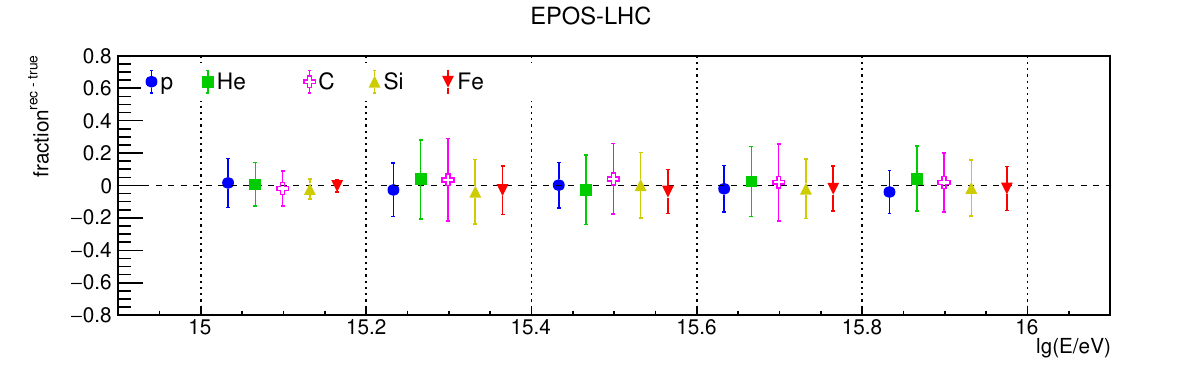}
    \caption{The values of $fraction^{rec - true}$ as a function of energy for the case in which the $LCm$ distributions are constructed on the basis of EPOS-LHC hadronic interaction model. The error bars represent the 68\% confidence contour and are considered as $\sigma^{2}_{syst}$ when fitting the experimental distributions.}
  \label{bias_syst_epos}
\end{figure}

Similar to the previously described procedure, we generated $LCm$ distributions based on a specific hadronic interaction model, considering random concentrations of the five primary particle species (p, He, C, Si, and Fe) within each energy interval considering the actual statistics of the data. These distributions encompassed all possible combinations. Subsequently, we fitted these distributions with MC predictions from the same model. We repeated this process 10000 times and, based on the fitting results, we constructed the $fraction^{rec-true}$ distributions for each species within each energy interval. From these distributions, we extracted the 68\% confidence contours, which we regarded as a second source of systematic errors denoted as $\sigma^{2}_{syst}$. Figure \ref{bias_syst_epos} presents the results for EPOS-LHC hadronic interaction model. The bias values for all species were found to be extremely close to zero, while the values of $\sigma^{2}_{syst}$ were smaller than $\sigma^{1}_{syst}$, as expected. Both $\sigma^{1}_{syst}$ and $\sigma^{2}_{syst}$ are applied to experimental results as separately error bars.

\section{Mass composition around the \textit{knee}}
\label{secMassComp}

In this study, we utilized KASCADE data to construct experimental $LCm$ distributions and compared them with MC predictions for five primary species (p, He, C, Si, and Fe) using three different hadronic interaction models (QGSjet-II-04, EPOS-LHC and SIBYLL 2.3d). We employed a Chi-Square minimization method to fit the experimental distributions, and we calculated the parameter errors using the Minos technique. 

\begin{figure}[tbp]
\centering 
\includegraphics[width=.49\textwidth]{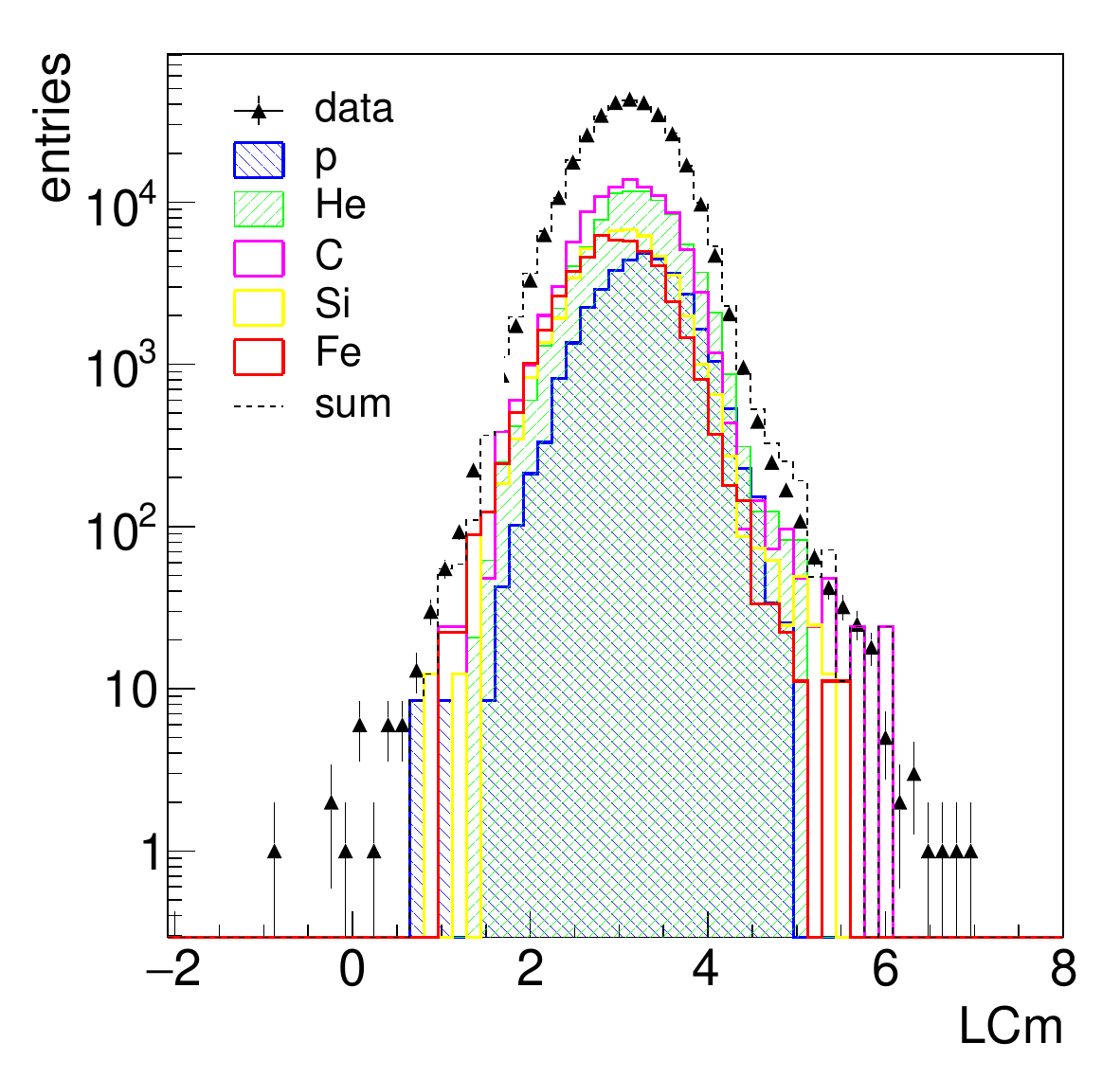}
\hfill
\includegraphics[width=.49\textwidth,origin=c]{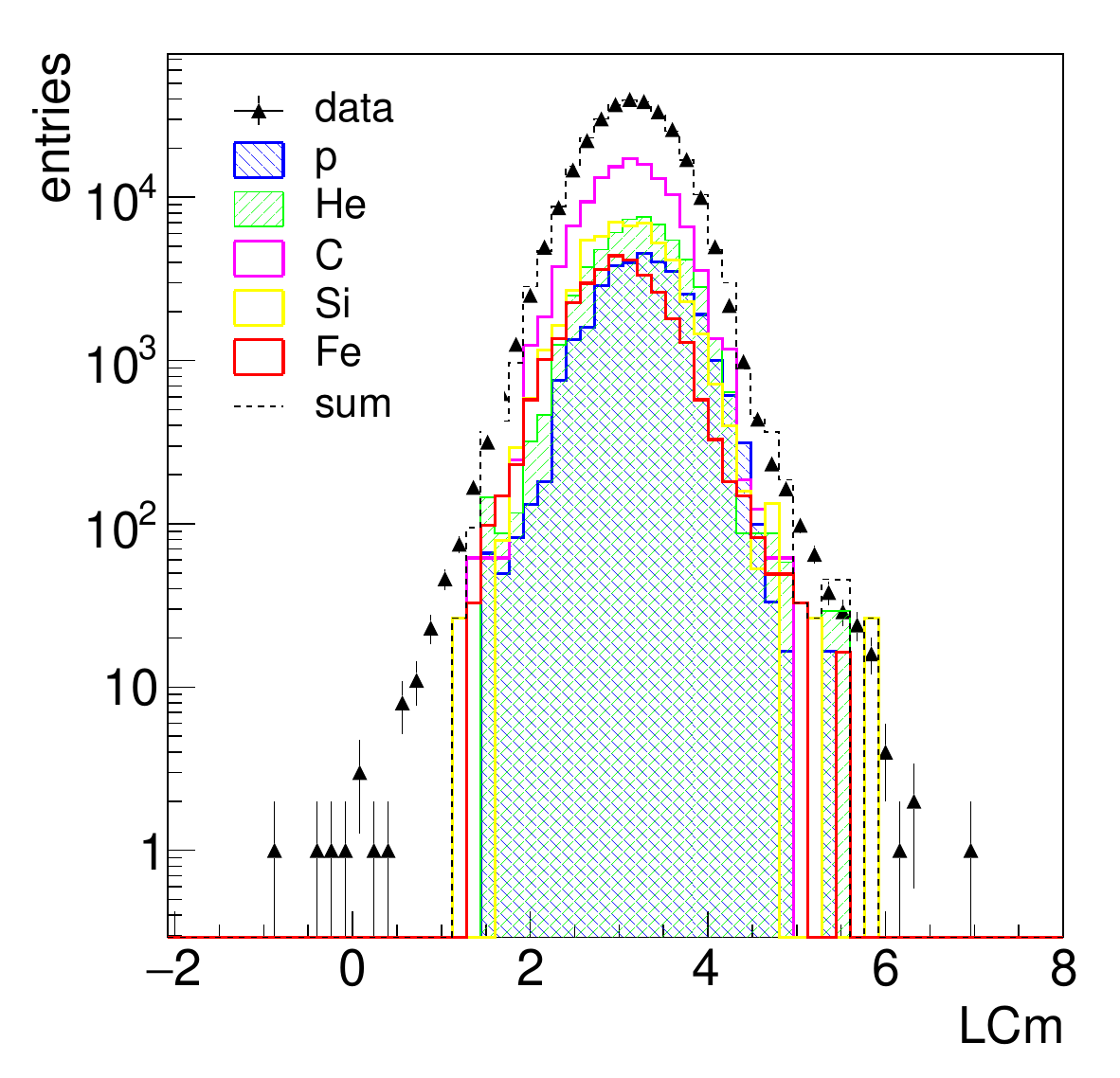}
\caption{\label{fig_fit} $LCm$ distribution reconstructed from KASCADE data in the energy interval $\lg(E/\rm eV) = [15.6 \text{ - } 15.8]$ \textit{(black triangles)}. The reconstructed fractions of each element (p, He, C, Si and Fe) are displayed on each plot. The \textit{(dashed line)} represents their sum as predicted by QGSjet-II-04 \textit{(left)} and EPOS-LHC \textit{(right)}.}
\end{figure}

In Figure \ref{fig_fit}, we present an example of an $LCm$ distribution reconstructed from KASCADE data in the energy interval $\lg(E/\rm eV) = [15.6 \text{ - } 15.8]$ (corrected for \textit{bin-to-bin migration} effect), along with the individual fractions of nuclei that best describe the observed distribution as predicted by the QGSjet-II-04 \textit{(left)} and EPOS-LHC \textit{(right)} hadronic interaction models. Since we have observed from simulations that the $LCm$ parameter shows minimal dependence on the choice of the hadronic interaction model, the obtained individual fractions of elements are in very good agreement. The lack of simulated distribution tails observed in Figure \ref{fig_fit}, resulting from small statistics compared to experimental data, cannot significantly influence the result of the fit.

To evaluate the fit quality in each energy interval we considered the Kolmogorov-Smirnov (KS) test which is a valuable tool for evaluating the goodness of fit in statistical modeling, particularly when the size of the data set is relatively large. This test is based on the comparison of the cumulative distribution function (CDF) of the fitted model with the observed data, which allows for a more comprehensive assessment of the agreement between the model and the data than other methods like the Chi-Squared test. Moreover, the KS test does not depend on the binning of the data or the model, making it more robust and reliable when working with limited statistics. A high KS probability indicates that the fitted model is a good match to the data, which means that the overall shape of the histograms of the two are very similar. The goodness-of-fit parameters (KS probability) evaluated for each energy interval and hadronic interaction model are listed in Table \ref{tab_1} and suggest, with few exceptions, an excellent agreement between data and all models on the entire energy range considered in this work.

\begin{table}[h]
\centering
\resizebox{0.9\textwidth}{!}{
\begin{tabular}{|l|*{5}{c|}}
\hline
$\lg (E/\rm eV)$ & $[15.0-15.2]$ & $[15.2-15.4]$ & $[15.4-15.6]$ & $[15.6-15.8]$ & $[15.8-16.0]$ \\
\hline
EPOS-LHC & 0.90 & 0.53 & 0.10 & 0.98 & 0.53  \\
QGSjet-II-04 & $< 0.1$ & 0.73 & 0.47 & 0.85 & 0.21  \\
SIBYLL 2.3d & $< 0.1$ & 0.18 & 0.39 & 0.98 & 0.54  \\
\hline
\end{tabular}
}
\caption{The KS probability as goodness-of-fit obtained after the comparison of the data with the fitted model.}
\label{tab_1}
\end{table}

\begin{figure}[tbp]
\centering 
\includegraphics[width=1.0\textwidth]{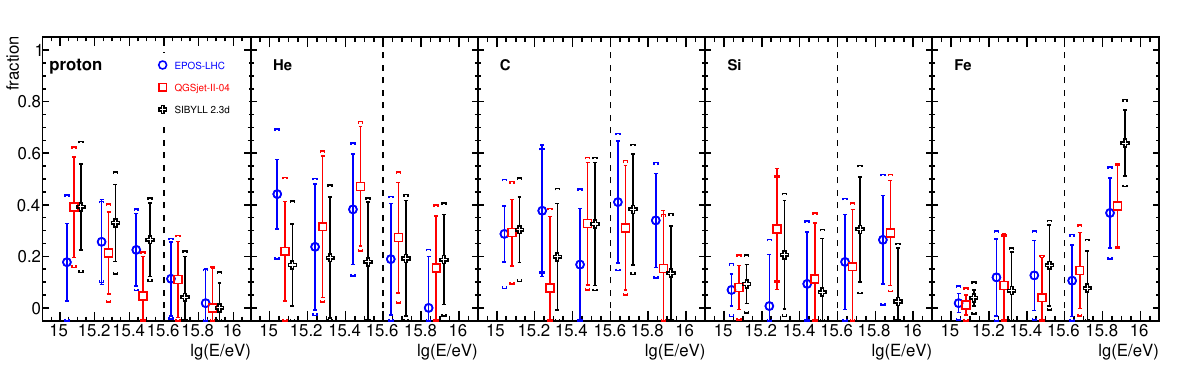}
\caption{\label{fig_fractions} The evolution of individual fractions of nuclei (p, He, C, Si and Fe) as a function of primary energy in the range $\lg(E/\rm eV) = [15.0 \text{ - } 16.0]$ as predicted by all three hadronic interaction models considered in this work. The two sets of systematic uncertainties are separately displayed. The parameter errors are typically smaller than the points size.}
\end{figure}

The evolution of the individual fractions of nuclei as a function of primary energy in the range $\lg(E/\rm eV) = [15.0 \textit{ - } 16.0]$ reconstructed on the basis of $LCm$ observable is represented in Figure \ref{fig_fractions} for all three hadronic interaction models considered.   

\begin{figure}[tbp]
\centering 
\includegraphics[width=1.0\textwidth]{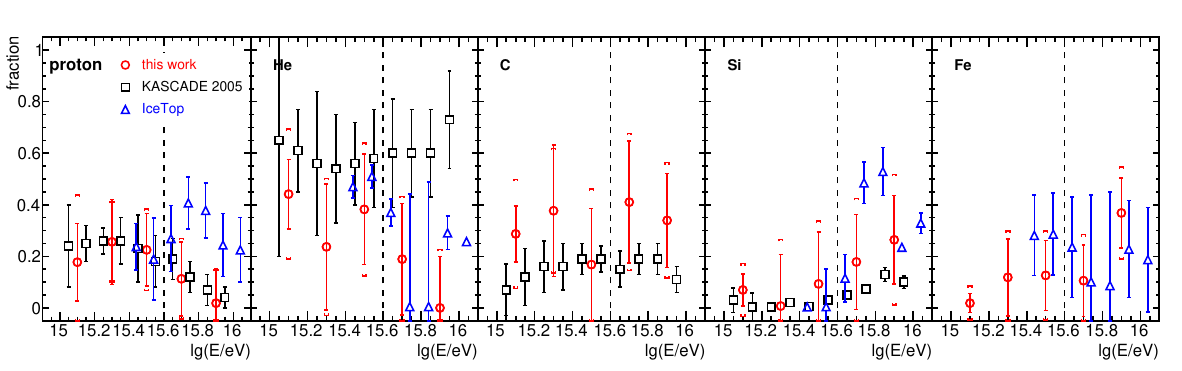}
\caption{\label{fig_icetop} The evolution of individual fractions of nuclei (p, He, C, Si and Fe) as a function of primary energy obtained in this work on the basis of EPOS-LHC model in comparison with the results obtained by KASCADE \cite{KASCADE:2005ynk} and IceTop \cite{Rawlins:2016bkc} experiments.}
\end{figure}

\begin{figure}[tbp]
\centering 
\includegraphics[width=1.0\textwidth]{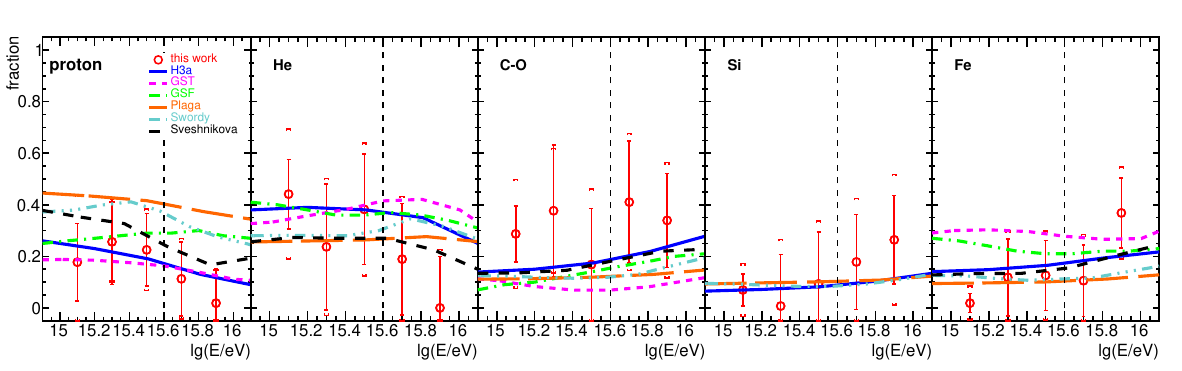}
\caption{\label{fig_allmodels} The evolution of individual fractions of nuclei (p, He, C, Si and Fe) as a function of primary energy obtained in this work on the basis of QGSjet-II-04 model in comparison with predictions of different astrophysical models: H3a \cite{Gaisser:2011klf}, GST \cite{Gaisser:2013bla}, GSF \cite{Dembinski:2017zsh}, Plaga \cite{Plaga:2001dk}, Swordy \cite{Swordy95} and Sveshnikova \cite{Sveshnikova:2003sa}.}
\end{figure}

An essential observation to highlight in these results is the remarkable agreement among all three hadronic interaction models regarding the individual abundance of each primary species as a function of energy. Additionally, the evolution of the individual fractions of nuclei with energy is highly consistent with previous studies that relied on different observables obtained from extensive air showers \cite{KASCADE:2004vrv, Apel:2008cd}. In Figure \ref{fig_icetop} we compare the results obtained in this work with those obtained by the KASCADE \cite{KASCADE:2005ynk} and IceTop \cite{Rawlins:2016bkc} experiments. We can assert that the results obtained through the method proposed in this work are in very good agreement with the results obtained from the KASCADE and IceTop experiments using different observables and techniques. Despite lacking important statistical significance, it is noteworthy that the fractions of heavy masses reconstructed from the data of the KASCADE experiment using this method are in complete agreement with the fractions reconstructed by the IceTop experiment.

These findings underscore the discriminative potential of the $LCm$ parameter for determining the mass composition of cosmic rays in experiments that use a relatively compact array of detectors e.g. \cite{DISCIASCIO2016166}.

In Figure \ref{fig_allmodels} we compare the mass composition obtained in this analysis with different astrophysical models that attempt to explain the origin of the \textit{knee} in the energy spectrum as a consequence of different acceleration and propagation scenarios \cite{Plaga:2001dk, Sveshnikova:2003sa, Swordy95}. We also performed a comparison with three more recent data-driven astrophysical models \cite{Gaisser:2011klf, Gaisser:2013bla, Dembinski:2017zsh}.

The astrophysical model of the \textit{knee} proposed by Plaga \cite{Plaga:2001dk} considers that cosmic rays are accelerated by "cannonballs" ejected in supernova explosions. The model proposed by Sveshnikova in \cite{Sveshnikova:2003sa} relates the presence of the \textit{knee} to the maximum energy attained by cosmic rays in shock fronts of supernova explosions, while the model proposed by Swordy \cite{Swordy95} assumes that the \textit{knee} is related to the leakage of cosmic rays from the Galaxy during the propagation process.

The mass composition around the \textit{knee}, obtained from KASCADE data using the $LCm$ parameter, displays a substantial level of agreement with all six astrophysical models regarding the evolutionary trend of primary abundance of different particle types as a function of energy. Notably, the H3a model \cite{Gaisser:2011klf} show a remarkable agreement with the relative abundances of individual species perfectly matching those derived from our analysis.

Based on the experimental capabilities of the LHAASO square kilometer array (KM2A) experiment \cite{LHAASO:2015pei,LHAASO:2022lxa}, which includes 5249 electromagnetic detectors spaced at 15 meters and 1188 muon detectors spaced at 30 meters, we can expect that the method proposed in this study is a suitable tool for reconstructing the mass composition around the knee from the forthcoming measurements of the LHAASO experiment.

\section{Summary and Conclusions}
\label{secConcl}

In this study we investigate the feasibility of discriminating the mass composition of cosmic rays at PeV energies by utilizing the $LCm$ parameter, which measures the signal non-uniformity recorded in detectors placed at a specific distance from the shower core at ground level.

Our analysis based on MC simulations using the KASCADE array revealed that the $LCm$ parameter is minimally dependent on the specific hadronic interaction model at high energies considered: QGSjet-II-04, EPOS-LHC and SIBYLL 2.3d.

Furthermore, we performed a mass composition analysis using data collected by the KASCADE experiment in the energy range $\lg(E/\rm eV) = [15.0 \text{ - } 16.0]$ around the \textit{knee}, with the $LCm$ parameter obtained from MC simulations for five primary species (p, He, C, Si, and Fe). Our results indicated a decrease in the abundance of protons and He primaries with increasing energy, while the heavier components became dominant above the \textit{knee}. The mass composition as a function of energy obtained in this work is in a good agreement with previous results reported by KASCADE and IceTop experiments. 

We also found that our results qualitatively support different astrophysical models that relate the origin of the \textit{knee} in the energy spectrum with different acceleration and propagation scenarios \cite{Plaga:2001dk,Swordy95,Sveshnikova:2003sa} while the  observed evolutionary trend of each primary abundance of the various types of particles with respect to energy is highly consistent with the predictions of three data-driven astrophysical models H3a \cite{Gaisser:2011klf}, GST \cite{Gaisser:2013bla} and GSF \cite{Dembinski:2017zsh}.

Based on these results, we conclude that the $LCm$ parameter can effectively serve as a minimally model-dependent discriminator for mass composition studies at PeV energies in experiments that use a relatively compact array of detectors. 

\acknowledgments
I am grateful to the KCDC team who made possible the access to data and simulations of the KASCADE experiment, especially to J\"{u}rgen Wochele for the assistance provided in accessing the data. 
I would like to thank Octavian Sima and the unknown referee for their useful comments and advice, as well as Ionel Lazanu and Mihaela Parvu for carefully reading the manuscript and providing valuable feedback.

This research was supported by Romanian Ministry of Research,
Innovation and Digitalization under Romanian National Core Program LAPLAS VII - contract no. 30N/2023.

\end{document}